\documentclass[12pt]{article}
 \usepackage{amsmath}
 \usepackage{amssymb}
 \usepackage{epsfig}
 
\begin{document}

\title{Calculation of Rare Decay $B^+  \to D_s^+ \overline K^{*0}$  \\
in Perturbative QCD Approach}
\author{Ying Li$^{a,b}$\footnote{e-mail: liying@mail.ihep.ac.cn}
  and Cai-Dian L\"u$^{a}$\footnote{e-mail: lucd@ihep.ac.cn}\\
{\it \small $a$  Institute of High Energy Physics,
CAS, P.O.Box 918(4)
  Beijing 100039, China}\\
{\it \small $b$ Physics Department, NanJing Normal University,
JiangSu 210097,China}
}

\maketitle
\begin{abstract}
The rare decay $B^+  \to D_s^+ \overline K^{*0}$ can occur only
via annihilation diagrams in the Standard Model. We calculate the
branching ratio in perturbative QCD approach based on  $k_T$
factorization theorem. We found that the branching ratio of this decay is
about of order $10^{-8}$, which may be sensitive to new physics.
\end{abstract}

  \begin{picture}(0,0)(0,0)
  \put(340,420){BIHEP-TH-2003-19}
  \end{picture}

\section{Introduction}
Rare $B$ decays are very important in particle physics,  because
they are important windows in testing the standard model (SM) and
they are sensitive to new physics. As a rather simple method,
factorization approach is accepted, because it explained
many decay branching ratios successfully \cite{bsw}. Recently,
some efforts have been made to improve their theoretical
application. One of these methods is the perturbative QCD approach
(PQCD),  in which many rare branching ratios such as $B \to K\pi$
\cite{lhs1}, $B \to \pi\pi$ \cite{pipi}, $B \to
\pi\rho$ \cite{pirho} were predicted.

In recent calculations, $B\to D_s^{(*)-}K^{(*)+}$, $B^+ \to
D_s^{(*)-} \overline K^0$ have been analyzed in the PQCD
approach \cite{lu:bdsk,ch}, leaving $B^+ \to D_s^+ \overline
K^{*0}$ not calculated. In decay $B^+ \to D_s^+ \overline K^{*0}$,
none of quarks in the final states is the same as one of the $B$ meson. This decay
is a pure annihilation type decay, which is described as $B$ meson
annihilating into vacuum and $D_s^+$ and $\overline K^{*0}$
produced from vacuum afterwards. Some information about PQCD picture can be
get in \cite{pqcd}.

 In PQCD, the decay amplitude is usually separated into
soft($\Phi$), hard($H$) and harder($C$) dynamics by different scales.
the factorization theorem allows us to write the decay amplitude as convolution,
\begin{equation}
 \mbox{Amplitude}
\sim \int\!\! d^4k_1 d^4k_2 d^4k_3\ \mathrm{Tr} \bigl[ C(t)
\Phi_B(k_1) \Phi_{D_s}(k_2) \Phi_{K^*}(k_3) H(k_1,k_2,k_3, t)
\bigr]. \label{eq:convolution1}
\end{equation}
In the function, $k_i(i=1,2,3)$ are momenta of light quarks in
each meson. $C(t)$ is Wilson coefficient which comes from the
QCD radiative corrections to the four quark operators.  $\Phi_M$ is the wave
function which describes the inner information of meson $M$. $H$
describes the four-quark operator and the quark pair from the sea
connected by a hard gluon whose scale is at the order of $M_B$, so
the $H$ can be perturbatively calculated. The hard part $H$ is
channel dependent, while $\Phi_M$ is independent of the specific
processes.

Some  analytic formulas for the decay amplitudes of $B^+  \to
D_s^+\overline K^{*0}$ decays  will be given in the next section.
In section \ref{sc:neval}, we give the numerical results and
discussion. Finally, we conclude this study in section
\ref{sc:concl}.

\section{$B^+  \to D_s^+ \overline K^{*0}$ amplitudes}\label{sc:formula}

For simplicity, we consider $B$ meson at rest. In the light-cone
coordinate, the $B$ meson momentum $P_1$, $D_s^+ $ momentum $P_2$
and $\overline K^{*0}$ momentum $P_3$ are taken to be:
\begin{equation}
       P_1 = \frac{M_B}{\sqrt{2}} (1,1,{\bf 0}_T),\ P_2 =
       \frac{M_B}{\sqrt{2}} (1-r_3^2,r_2^2,{\bf 0}_T),\ P_3 =
       \frac{M_B}{\sqrt{2}} (r_3^2,1-r_2^2,{\bf 0}_T) , \label{eq:momentun1}
\end{equation}
where $r_2= M_{D_s^+}/M_B$ and $r_3=M_{\overline K^{*0}}/M_B$.
Putting the light (anti-)quark momenta in $B^+$, $D_s^+$ and
$\overline K^{*0}$ mesons as $k_1$, $k_2$, and $k_3$,
respectively, we can choose
\begin{equation}
k_1 = (x_1P_1^+, 0, {\bf k}_{1T}),\ k_2 = (x_2 P_2^+, 0, {\bf
k}_{2T}), \  k_3 = (0, x_3 P_3^-, {\bf k}_{3T}) .
\label{eq:momentun2}
\end{equation}
The $\overline K^{*0}$ meson's longitudinal polarization vector
$\epsilon$ and  transverse polarization vector $\epsilon_T$ are
given by :
\begin{equation}
\epsilon_L= \frac{M_B}{\sqrt{2}M_{\overline K^{*0}}} (-r_3^2,
1-r_2^2, {\bf 0}_T)  ,  \  ~~ \epsilon_T
=\frac{M_B}{\sqrt{2}M_{\overline K^{*0}}} (0, 0, {\bf 1}_T).
\label{eq:pola}
\end{equation}
Then, integration over $k_1^-$, $k_2^-$, and $k_3^+$ in
eq.(\ref{eq:convolution1}) leads to:
\begin{multline}
 \mbox{Amplitude}
\sim \int\!\! d x_1 d x_2 d x_3
b_1 d b_1 b_2 d b_2 b_3 d b_3 \\
\mathrm{Tr} \bigl[ C(t) \Phi_B(x_1,b_1)
\Phi_{D_s}(x_2,b_2,\epsilon) \Phi_{K^*}(x_3, b_3) H(x_i,
b_i,\epsilon, t) S_t(x_i)\, e^{-S(t)} \bigr],
\label{eq:convolution2}
\end{multline}
where $b_i$ is the conjugate space coordinate of $k_{iT}$, and $t$
is the largest energy scale in $H$, as the function in terms of
$x_i$ and $b_i$. The large logarithms ($\ln m_W/t$) coming from
QCD radiative corrections to four quark operators are included in
the Wilson coefficients $C(t)$. The large double logarithms
($\ln^2 x_i$) on the longitudinal direction are summed by the
threshold resummation \cite{L3}, and they lead to a jet function  $S_t(x_i)$ which
smears the end-point singularities on $x_i$. The last term,
$e^{-S(t)}$, contains two kinds of logarithms. One of the large
logarithms is due to the renormalization of ultra-violet
divergence $\ln tb$, the other is double logarithm $\ln^2 b$ from
the overlap of collinear and soft gluon corrections. This Sudakov
form factor suppresses the soft dynamics effectively \cite{soft}.
Thus it makes perturbative calculation of the hard part $H$
reliable.

The heavy $B$ and $D_s$ meson wave functions are restricted by heavy quark
symmetry. In the heavy quark limit, we may use only one independent distribution amplitude
for each of them \cite{ly}.
\begin{equation}
 \Phi_{M}(x,b) = \frac{i}{\sqrt{6}}
\left[ (\not \! P_1 \gamma_5) + M \gamma_{5} \right]
\phi_M(x,b),
\end{equation}
 where $M=B$, $D_s$. For the light  $K^*$ meson, only the
 longitudinal wave function for outgoing state is relevant, which
 is written as:
\begin{equation}
 \Phi_{K^*}(x_3,b_3) = \frac{i}{\sqrt{6}} \Bigl[ M_{K^*} \not \!
 \epsilon_L
\phi_{K^*}(x_3,b_3) +\not \! \epsilon_L \not \!P_3
\phi_{K^*}^t(x_3,b_3) + M_{K^*} I \phi_{K^*}^s(x_3,b_3) \Bigr].
\end{equation}
Unlike the heavy mesons, there are three different distribution amplitudes.

\begin{figure}[htb]
     \epsfig{file=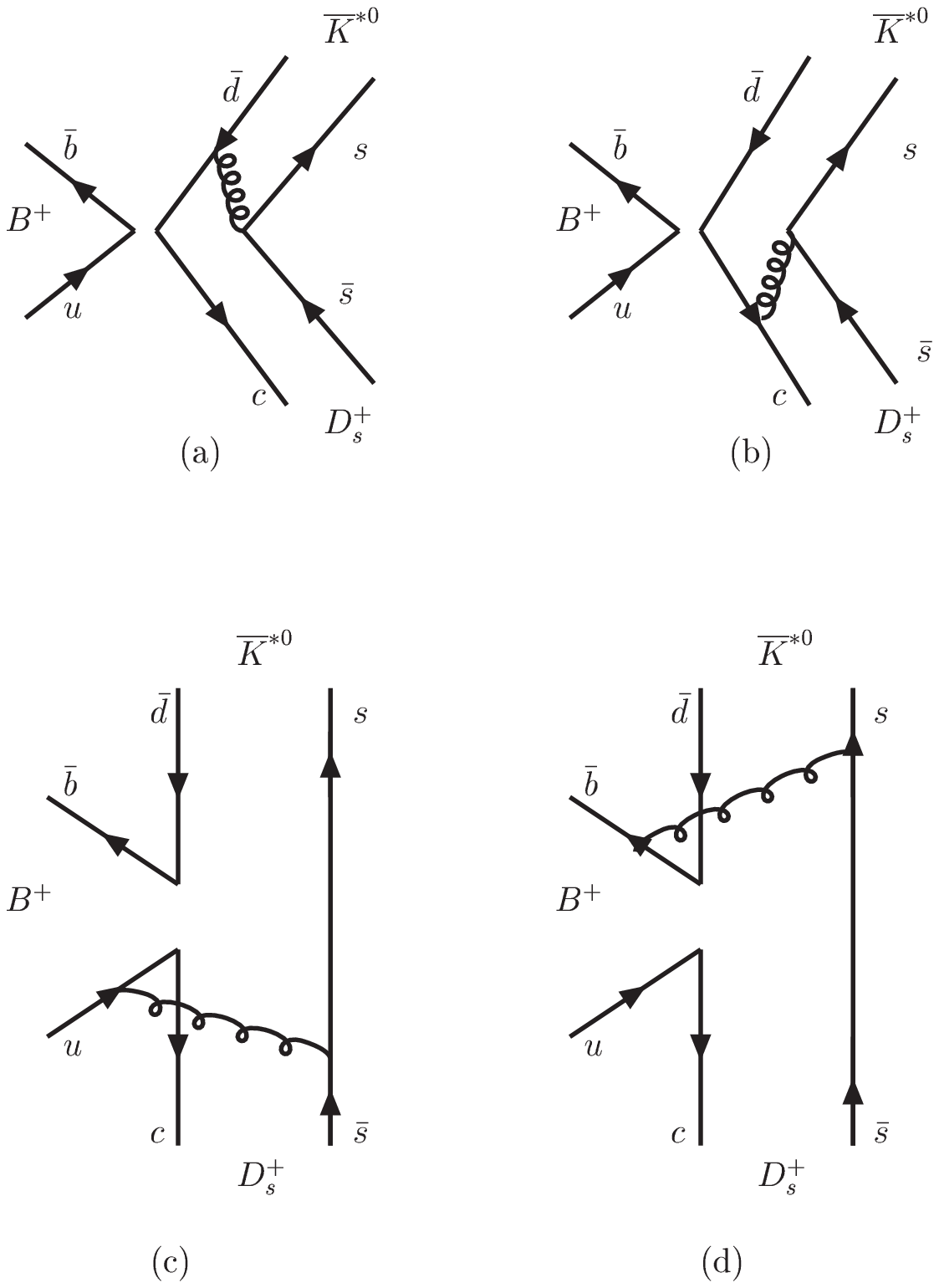,bbllx=.1cm,bblly=10.4cm,bburx=4cm,bbury=25cm,%
 width=2.6cm,angle=0}
  \caption{Diagrams for $B^+ \to D_s^+\overline K^{*0}$ decay. The factorizable
  diagrams (a),(b) contribute to $F_a$, and nonfactorizable (c),
  (d) do to $M_a$.}
  \label{fig:diagrams1}
 \end{figure}

In the decay $B^+  \to D_s^+ \overline K^{*0}$, the effective
Hamiltonian at the scale lower than $M_W$ is the same as $B^+ \to
D_s^+ \overline K^0$ decay
\begin{gather}
 H_\mathrm{eff} = \frac{G_F}{\sqrt{2}} V_{ub}^*V_{cd} \left[
C_1(\mu) O_1(\mu) + C_2(\mu) O_2(\mu) \right], \\
  O_1 = (\bar{b}\gamma_\mu P_L d) (\bar{c}\gamma^\mu P_Lu), \quad
  O_2 = (\bar{b}\gamma_\mu P_Lu) (\bar{c}\gamma^\mu P_Ld),
\end{gather}
where the projection operator is defined by $P_L=(1-\gamma_5)/ 2
$. $V_{ub}^*V_{cd}$ are the products of the CKM matrix elements,
and $C_{1,2}(\mu)$ are the Wilson coefficients. According to
the effective Hamiltonian, the lowest order diagrams of $B^+  \to
D_s^+ \overline K^{*0}$ are drawn in Fig.\ref{fig:diagrams1}. As
stated above, this decay only has annihilation diagrams.

After perturbative QCD calculations, we get  the
 factorizable contribution $F_a$, and the nonfactorizable contribution
 $M_a$
 illustrated by Fig.1(a)(b)
and Fig.1(c)(d), respectively.
\begin{multline}
F_a = 16\pi C_F M_B^2 \int_0^1\!\!\! dx_2 dx_3
 \int_0^\infty\!\!\!\!\!  b_2 db_2\, b_3 db_3\ \phi_{D_s}(x_2,b_2) \\
\times \Bigl[ \bigl\{
(1-x_3-3r_2^2+2x_3r_2^2) \phi_{K^*}(x_3,b_3)
+ r_2 \left( 1-2x_3 \right) r_3 \phi_{K^*}^t(x_3,b_3) \\
- r_2 (3-2x_3) r_3 \phi_{K^*}^s(x_3,b_3)
\bigr\} E_{f}(t_a^1) h_a(x_2,x_3,b_2,b_3) \\
- \bigl\{
(1-r_2^2)x_2\phi_{K^*}(x_3,b_3) \\
- 2r_2 (1+x_2) r_3 \phi_{K^*}^s(x_3,b_3) \bigr\} E_{f}(t_a^2)
h_a(1-x_3,1-x_2,b_3,b_2) \Bigr], \label{eq:Fa}
\end{multline}
\begin{multline}
M_a =  \frac{1}{\sqrt{2N_c}} 64\pi C_F M_B^2 \int_0^1\!\!\! dx_1
dx_2 dx_3
 \int_0^\infty\!\!\!\!\! b_1 db_1\, b_2 db_2\
 \phi_B(x_1,b_1) \phi_{D_s}(x_2,b_2) \\
\times \Bigl[ \bigl\{ x_2(1-2r_2^2) \phi_{K^*}(x_3,b_2)
 + r_2 \left(x_2+x_3-1\right) r_3 \phi_{K^*}^t(x_3,b_2) \\
 - r_2 \left(x_2-x_3+1\right) r_3 \phi_{K^*}^s(x_3,b_2)
\bigr\}
E_{m}'(t_{m}^1) h_a^{(1)}(x_1, x_2,x_3,b_1,b_2) \\
- \bigl\{
 \left( -(x_2+2x_3-2)r_2^2+x_3 \right) \phi_{K^*}(x_3,b_2)
 + r_2 \left(1-x_2-x_3 \right) r_3 \phi_{K^*}^t(x_3,b_2) \\
 - r_2 \left(3+x_2-x_3 \right) r_3 \phi_{K^*}^s(x_3,b_2)
\bigr\}E_{m}'(t_{m}^2) h_a^{(2)}(x_1, x_2,x_3,b_1,b_2) \Bigr].
\label{eq:Ma}
\end{multline}
In our work, $r_3^2$ , $ r_2^2$ and $x_1$
in numerators are neglected. In the functions, $C_F = 4/3$ is the
group factor of $\mathrm{SU}(3)_c$ gauge group, and the functions
$E_{f}$, $t_a^{1,2}$, $h_a$ are given in the appendix of
\cite{lu:bdsk}.

Comparing with the previous calculated  $B^0 \to D_s^{(*)-} K^+$ decay \cite{lu:bdsk},
we find that
the leading twist contribution, which is proportional to
$\phi_K^A$, is almost the same. However the sub-leading twist
contribution, which is proportional to $r_2r_3$ in (\ref{eq:Fa})
and (\ref{eq:Ma}), is quite different.

The total decay width  for $B^+  \to D_s^+ \overline K^{*0}$
decay is given as
\begin{equation}
 \Gamma(B^+ \to D_s^+ \overline K^{*0}) = \frac{G_F^2 M_B^3}{128\pi} (1-r_2^2)
|V_{ub}^*V_{cd} ( f_B F_a+ M_a)|^2. \label{eq:neut_width}
\end{equation}
The decay width for $CP$ conjugated model, $B^- \to D_s^-
\overline K^{*0}$, is the same value as $B^+  \to D_s^+ \overline
K^{*0}$, just replace $V_{ub}^*V_{cd}$ with $V_{ub}V_{cd}^*$.
Since there is only one kind of CKM phase involved in this decay,
there is no CP violation in the standard model.

\section{Numerical Results}\label{sc:neval}

We use the same $B$ and $D_s$ meson wave functions  as   before
\cite{lu:bdsk}
\begin{equation}
\phi_B(x,b) = N_B x^2(1-x)^2 \exp \left[
-\frac{M_B^2\ x^2}{2 \omega_b^2} -\frac{1}{2} (\omega_b b)^2
\right],
\end{equation}
\begin{equation}
\phi_{D_s}(x,b) = \frac{3}{\sqrt{2 N_c}} f_{D_s} x(1-x)\{ 1 +
a_{D_s} (1 -2x) \}\exp \left[-\frac{1}{2} (\omega_{D_s} b)^2
\right].
\end{equation}
The $K^*$ meson's distribution amplitudes  are given by light cone QCD sum rules
\cite{Ball:1998je}:
\begin{eqnarray}
\phi_{K^*}(x)& =&\frac{f_{K^*}}{2\sqrt{2 N_c}} 6 x(1-x) \left\{1
+ 0.57\cdot (1-2x) + 0.07 \cdot C_2^{3/2}(1-2x) \right\}, \\
\phi_{K^*}^t(x)& =& \frac{f_{K^*}^T}{2\sqrt{2 N_c}}
\left\{0.3(1-2x)[3(1-2x)^2+10(1-2x)-1]+1.68 C_4^{1/2}(1-2x) \right.  \nonumber \\
& &\left.+0.06(1-2x)^2[5(1-2x)^2-3]+0.36[1-2(1-2x)(1+\ln(1-x)]\right\}, \\
\phi_{K^*}^s(x) & = &\frac{f_{K^*}^T}{2\sqrt{2 N_c}}\left\{3
(1-2x) [1+0.2 (1-2x)+0.6 (10x^2-10x+1)] \right.\nonumber \\
& &\left. -0.12x(1-x)+0.36[1-6x-2\ln(1-x)]\right\},
\end{eqnarray}
with the Gegenbauer polynomials,
\begin{equation}
C_2^{1/2}(\xi)=\frac{1}{2}(3\xi^2-1),   ~~~
C_4^{1/2}(\xi)=\frac{1}{8}(35\xi^4-30\xi^2+3).
\end{equation}
In addition, we use the following input parameters, for meson decay constants and
 the  CKM matrix elements and the lifetime of
$B^+$ \cite{pdg},
\begin{gather}
f_B = 190 \mbox{ MeV},\ ~~ f_{D_s} = 220 \mbox{ MeV},\  ~~ f_{K^*}^{(T)} =
200 \mbox{ MeV},    \\
|V_{ub}|= 3.6 \times 10^{-3}, ~~  \  |V_{cd}|=0.224,   \  ~~
  \tau_{B^+}=1.67\times 10^{-12}\mbox{ s.}
\end{gather}

The branching ratio obtained from the analytic formulas may be
sensitive to many parameters especially those in the meson wave
functions \cite{lu:bdsk,ch}.
Similar to the $B \to D_S^{(*)}K$ decays \cite{lu:bdsk,ch}, we found that
the branching ratio is
not sensitive to the parameter change of the $K^*$ meson wave
function.  but they do be sensitive to the heavy B and $D_s$ meson
wave function parameters.
For illustration of the uncertainties of the branching ratios, we
choose the following $B$ and $D_s$ meson wave function parameters
\begin{align}
  0.35  \mathrm{GeV} \leq  \omega_b  \leq 0.45  \mathrm{GeV},\\
 0.21  \mathrm{GeV} \leq  \omega_{D_s}  \leq 0.30  \mathrm{GeV} ,\\
 0.21  \mathrm{GeV} \leq  a_{D_s}  \leq 0.30  \mathrm{GeV}.
\end{align}
Using the range stated above, the branching ratio
normalized by the decay constants and the CKM matrix elements
results in :
\begin{gather}
\mathrm{Br}(B^+ \to D_s^+ \overline{K}^{*0}) = (1.8 \pm{0.3})
\times 10^{-8} \left( \frac{f_B\ f_{D_s}}{190\mbox{ MeV}\cdot
240\mbox{ MeV}} \right)^2 \left( \frac{|V_{ub}^*\ V_{cd}|} {
0.0036 \cdot 0.224} \right)^2.
\end{gather}
Considering the uncertainty of $f_B$, $f_{D_s}$ and the CKM matrix
elements, the branching ratio of the $B^+ \to D_s^+
\overline{K}^{*0}$ decay is at the order of $10^{-8}$. This is
still far from the current experimental upper limit \cite{pdg},
\begin{align}
\mathrm{Br}(B^+ \to D_s^+ \overline{K}^{*0}) & < 5 \times
10^{-4}.\label{eq:brex2}
\end{align}

\begin{table}
\caption{Decay branching ratios calculated in PQCD approach using the same
parameters for $B$ and $D_s$ wave functions, and $f_{D_s}
=f_{D_s^*} = 220$MeV.}
\begin{tabular}{cccc}
\hline \hline
Decay channel &  $D_s^+ \bar K^0$  & $D_s^{*+} \bar K^0$  & $D_s^+ \bar K^{*0}$
\\
Br ( $10^{-8}$ )   &      $1.5\pm 0.2 $    &   $3.7\pm0.5$      &
$1.8\pm 0.3 $\\
\hline   \hline
\end{tabular}
\end{table}

For comparison, we list the branching ratios of $B^+ \to D_s^{(*)+} \overline{K}^0$
and $B^+ \to D_s^{+} \overline{K}^{*0}$  calculated in PQCD
approach, using the same parameters for $B$ and $D_s^{(*)}$ wave
functions. For simplicity, we set $f_{D_s} =f_{D_s^*} =220 $MeV.
 From the above result, we can see that the branching ratio of
 $B^+ \to D_s^{+} \overline{K}^{*0}$ is  a little larger
than that of $B^+ \to D_s^{+} \overline{K}^0$ \cite{lu:bdsk}. From
equation (\ref{eq:Fa},\ref{eq:Ma}) and formulas in ref.\cite{lu:bdsk},
 we find the coefficients of
sub-leading twist become negative, but they are all proportional
to $r_2r_3$ which is a little bit  small.
 Because of the suppression of
CKM matrix elements, the branching ratio of $B^+ \to D_s^+
\overline{K}^{*0}$ is much smaller than that of the neutral decay $B^0 \to D_s^-
\overline{K}^{*+}$.

  \begin{figure}[htb]
        \epsfig{file=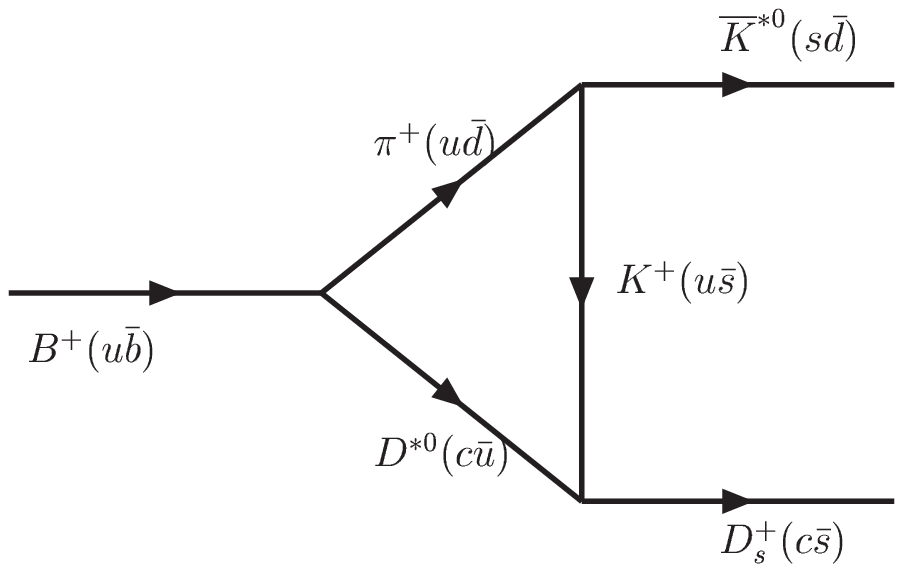,bbllx=3cm,bblly=20cm,bburx=6.4cm,bbury=26cm,%
    width=3cm,angle=0}
   \caption{Typical diagram for picture of final state interaction.}
   \label{fig:diagrams2}
  \end{figure}

Since pure annihilation type B decays are suppressed comparing to
spectator diagram decays, the soft final state interactions may be
important \cite{yang}. In our case,
$B^+$ meson  can decay into $D^{*0}$ and $\pi^+$ , the
secondary particles then exchanging a $K^+$, scatter into $D_s^+$,
$\overline K^{*0} $ through final state interaction afterwards.
This picture is depicted in Fig.\ref{fig:diagrams2}, which  is
difficult to
 calculate accurately, since final state interaction is purely non-perturbative
  \cite{yang}. In ref.\cite{lu:bdsk}, the
results from PQCD approach for $B^0 \to D_s^- K^+$ decay were
consistent with the experiment, which shows that the soft final
state interaction may not be important.

\section{Conclusion} \label{sc:concl}

In hadronic  two-body $B$ meson decays, the energy release is larger than
1 GeV. The final state mesons are
moving very fast.
  The soft final states interaction  may not be important in
the two-body $B$ meson decays.  The PQCD approach based on $k_T$
factorization theorem is applicable to the calculation of B meson
decays.

In this work, we calculate the $B^+ \to D_s^+ \overline{K}^{*0}$
decay in the PQCD approach. Since neither of quarks in $B^+$
appeared in the final state mesons, this process occurs only via
annihilation type diagrams. From our PQCD work, the branching
ratio of $B^+ \to D_s^+ \overline{K}^{*0}$ is small in SM, which
is around $10^{-8}$. This branching ratio will be measured in the
LHC-B in future. This small branching ratio predicted in the SM,
makes the channel sensitive to any new physics contribution.


                       \end{document}